# Semiconductive and Ferromagnetic Lanthanide MXenes Derived from Carbon Intercalated Two-dimensional Halides

Qian Fang[1,2,3†], Liming Wang[4†], Kai Chang[4], Hongxin Yang[4*], Pu Yan[5], Kecheng Cao[5], Mian Li[1,3], Zhifang Chai[1,3], Qing Huang[1,3*]

[1]*Zhejiang Key Laboratory of Data-Driven High-Safety Energy Materials and Applications, Ningbo Key Laboratory of Special Energy Materials and Chemistry*，*Ningbo Institute of Materials Technology and Engineering, Chinese Academy of Sciences, Ningbo, 315201, China*

[2]*University of Chinese Academy of Sciences, Beijing, 100049, China*

[3]*Qianwan Institute of CNiTECH, Ningbo, 315336, China*

[4]*Center for Quantum Matter, School of Physics, Zhejiang University, Hangzhou, 31008, China*

5*School of Physical Science and Technology & Shanghai Key Laboratory of High-resolution Electron Microscopy, ShanghaiTech University; Shanghai, 201210, China*

†*These authors contributed equally to this work.*

Email: huangqing@nimte.ac.cn, hongxin.yang@zju.edu.cn

**Abstract**

Two-dimensional (2D) magnetic semiconductors are a key focus in developing next-generation information storage technologies[1,2]. MXenes, as emerging 2D early transition metal carbides and nitrides, offer versatile compositions and tunable chemical structures[3]. Incorporating lanthanide metals, with their unique role of 4*f*-electrons in engineering physical properties[4], into MXenes holds potential for advancing technological applications. However, the scarcity of lanthanide-containing ternary MAX phase precursors and the propensity of lanthanides to oxidize pose significant challenges to obtain lanthanide MXenes ($Ln_2CT_2$) via the 'top-down' etching method[5]. Here, we propose a general 'bottom-up' methodology for lanthanide MXenes, that derive from carbon intercalated *van der Waals* building blocks of 2D halides. Compared to conventional MXenes conductors, the synthesized $Ln_2CT_2$ exhibit tunable band gaps spanning 0.32 eV to 1.22 eV that cover typical semiconductors such as Si (1.12 eV) and Ge (0.67 eV). Additionally, the presence of unpaired *f*-electrons endows $Ln_2CT_2$ with intrinsic ferromagnetism, with Curie temperatures ranging between 36 K and 60 K. Theoretical calculations reveal that, in contrast to traditional MXenes, the number of *d*-electrons states around the Fermi level are largely diminishes in bare $Ln_2C$ MXenes, and the halogen terminals can further exhaust these electrons to open band gaps. Meanwhile, the Ln-4*f* electrons in $Ln_2CT_2$ are highly localized and stay away from the Fermi level, contributing to the spin splitting for the observed ferromagnetic behavior. Lanthanide MXenes hold immense promise for revolutionizing future applications in spintronic devices.

**Introduction**

Two-dimensional (2D) early transition metal carbides and nitrides (MXenes) exhibit the diversity and tunability of chemical compositions, resulting in a family that exceeds 50 MXenes[6]. This characteristic renders them a promising material system for designing 2D magnetic semiconductors[7]. The general chemical formula of MXenes is $M_{n+1}X_nT_x$ ($n$=1-4), where M denotes an early transition metal (e.g., Ti, V, Nb), X represents carbon or nitrogen, and T signifies surface terminations, including -OH, chalcogens and halogens. To date, the majority of materials reported are non-magnetic conductors, including $Ti_3C_2T_x$ (where T represents -O, -OH and -F)[8]. Incorporating elements with magnetic moments, like lanthanide metal, into the M site is a crucial strategy for modulating the physical properties of MXenes.

MXenes are typically synthesized via a top-down strategy. This process begins with the layered MAX phases ($M_{n+1}AX_n$, where A normally represents a main group element) and employs fluorine-containing acids or molten salts to selectively etch the A element layers, thereby retaining the surface-terminated $M_{n+1}X_n$ layered structures[8-10]. However, the absence of ternary MAX phases with lanthanides at the M site ($Ln_{n+1}AX_n$) has hindered the synthesis of the corresponding MXenes ($Ln_{n+1}X_nT_x$). By employing an alloying method, a series of $i$-MAX phases with M site solid solutions, such as $(Mo_{2/3}Ln_{1/3})_2AlC$, have been synthesized, where Mo and lanthanide atoms are orderly arranged at the M site[5]. Unfortunately, the lanthanide atoms were removed along with the Al atoms during the etching process to synthesize MXenes, leaving behind $Mo_{1.33}C$ nanosheets with distinct ordered metal vacancies. An alternative approach is the bottom-up synthesis directly from chemical components, avoiding the etching step that would introduce defects and structural damage. The effectiveness of this concept has been demonstrated through chemical vapor deposition (CVD) technique, obtaining $Ti_2CCl_2$ and even unexploited $Ti_2NCl_2$, $Zr_2CCl_2$ MXenes[11]. Analogous structures of rare earth carbides, such as $Sc_2CCl_2$ and $Y_2CCl_2$, were claimed in the 1980s using solid-state reaction[12]. However, it remains unclear how these chemical reactions drive the transformation of these atoms into the layered structures of MXenes, and no attempt is reported on the lanthanide MXenes.

Here, we employ a bottom-up approach from 2D building blocks to synthesize lanthanide MXenes ($Ln_2CT_2$, Ln=Gd, Tb, Dy, Ho, Er, Lu; T=Cl, Br). Furthermore, the prepared $Ln_2CT_2$ MXenes possess both semiconducting and ferromagnetic properties that are uncommon in transition metal MXenes. This method not only broadens the range of M elements typically hard to preserve in the $M_{n+1}X_n$ structures via the traditional selective etching approach but also expands the potential applications of MXenes in the field of spintronics.

# Lanthanide MXenes derived from 2D lanthanide halides

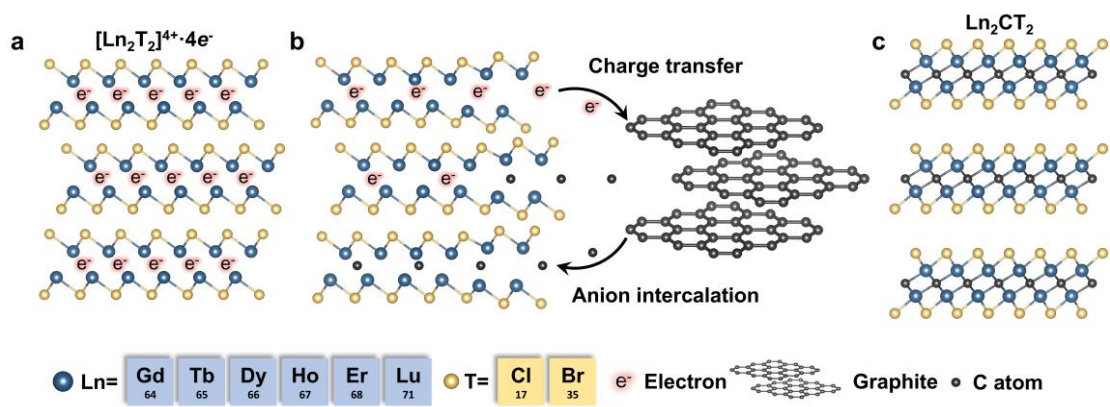

**Fig. 1 Schematic diagram of synthesis process of Ln$_2$CT$_2$.** (**a**) The structure of lanthanide mono-halides (LnT) obtained by the reaction between copper halides and lanthanide metals. (**b**) The transformation mechanism from LnT to Ln$_2$CT$_2$: The IAEs within the interlayer of LnT reduce graphite into carbon anions (C$^{4-}$), and C$^{4-}$ enters the octahedral interstitial positions within the double metal structure. (**c**) The structure of obtained Ln$_2$CT$_2$ (Ln=Gd, Tb, Dy, Ho, Er, Lu; T=Cl, Br).

Lanthanide mono-halides (LnT), which are a class of 2D building blocks with *van der waals* (*vdW*) gaps, serve as crucial starting precursors for the anion intercalation. The LnT crystal has the feature of 2D double-metal (Ln-Ln) frameworks arising from the condensation of [Ln$_6$] octahedra (**Fig. 1a**)[13]. Such double-metal layers are enveloped by halogen atoms (T) in a T-Ln-Ln-T stacking sequence, resembling the structures of halogen-terminated MXenes devoid of X atoms. During the synthesis, controlled molar ratios of copper halides (CuT$_2$) and lanthanide metals (Ln) are utilized to form LnT *in situ* (**Eq. 1** and **Fig S1**). LnT are also isostructural with the previously reported electrides of rare-earth mono-chlorides, [ReCl]$^{2+}$·2$e^-$ (Re=Y, La), containing interstitial anionic electrons (IAEs) within the Re-Re layers[14]. Given the similar chemical properties of rare earth elements, the formula of LnT can be written as [LnT]$^{2+}$·2$e^-$ or [Ln$_2$T$_2$]$^{4+}$·4$e^-$ to account for the double metal structure, which act as strong electron donors to reduce graphite (C) to carbon anions (C$^{4-}$) (**Eq. 2** and **Fig. 1b**). Concurrently, the as-formed C$^{4-}$ anions spontaneously diffuse into positively charged octahedral gaps of double metal halogenide [Ln$_2$T$_2$]$^{4+}$ driving by electrostatic force (**Eq. 3** and **Fig. 1b**). Lanthanide MXenes (Ln$_2$CT$_2$, Ln=Gd, Tb, Dy, Ho, Er, Lu; T=Cl, Br) (**Fig. 1c**) are successfully synthesized using this approach.

(1) 2Ln+CuT$_2$=2LnT+Cu 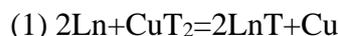

(2) [Ln$_2$T$_2$]$^{4+}$·4$e^-$+C=[Ln$_2$T$_2$]$^{4+}$+C$^{4-}$ 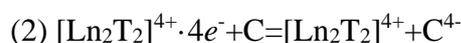

(3) [Ln$_2$T$_2$]$^{4+}$+C$^{4-}$=Ln$_2$CT$_2$ 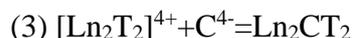

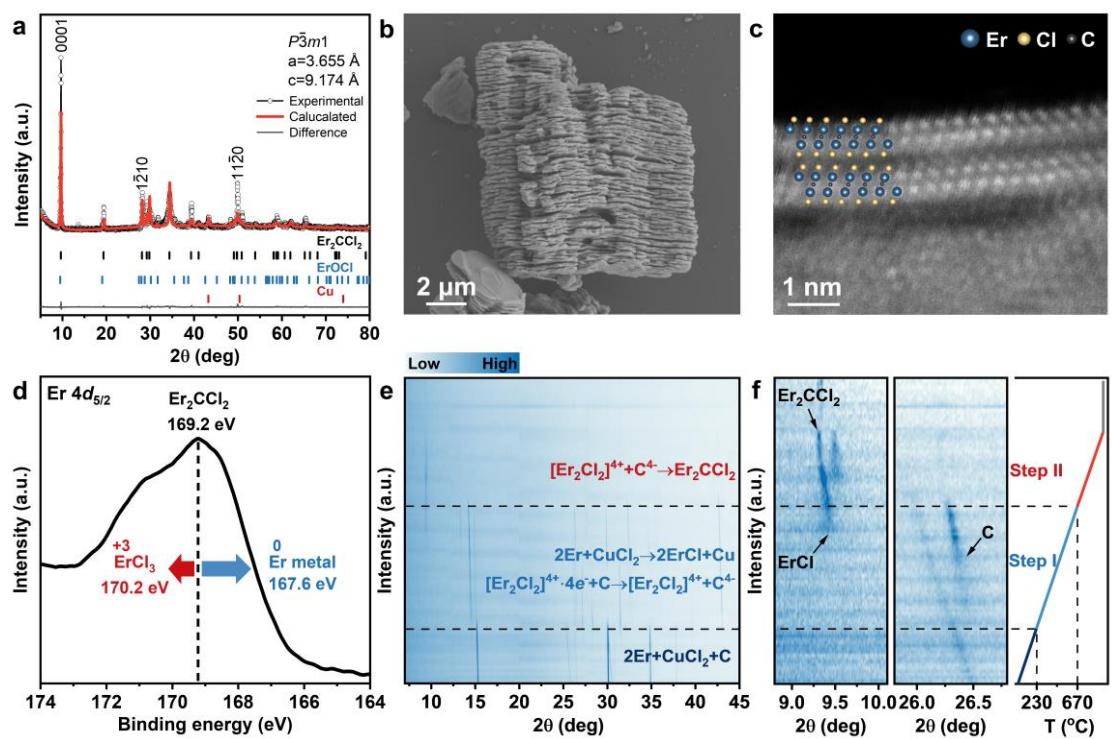

**Fig. 2 Structural and morphological characterization of Er$_2$CCl$_2$.** (**a**) XRD pattern and Rietveld refinement of Er$_2$CCl$_2$. (**b**) SEM image of Er$_2$CCl$_2$. (**c**) STEM image of Er$_2$CCl$_2$. (**d**) Er $4d_{5/2}$ orbital XPS spectra of Er$_2$CCl$_2$. (**e** and **f**) The heatmaps form of SR-XRD patterns as a function of temperature during the synthesis of Er$_2$CCl$_2$ ($\lambda$=1.54 Å). The corresponding temperature change curve is presented on the right, with different colors highlighting the distinct stages of the reaction. The enlarged patterns of $2\theta$=8.8 °-10.1 ° and 25.8 °-26.8 ° illustrate the reaction between ErCl and C, as well as the emergence of Er$_2$CCl$_2$.

To validate the proposed synthesis strategy, yttrium mono-chloride (YCl), one of confirmed electrides ([YCl]$^{2+}\cdot 2e^-$)[14], was synthesized *in situ* to serve as a 2D *vdW* building block. Electrons in YCl are captured by carbon, resulting in a positively charged double metal layer of yttrium ([Y$_2$Cl$_2$]$^{4+}$), which facilitates the intercalation of the reduced C$^{4-}$ into their interlayer space to form the Y$_2$CCl$_2$ MXenes. The X-ray diffraction (XRD) spectrum of the resulting product clearly indicated the formation of the Y$_2$CCl$_2$ phase, which shares an identical atomic arrangement with MXenes in their single-layer crystal structures[12] (**Fig. S2**). Using a similar approach, Sc$_2$CCl$_2$ MXenes was also successfully synthesized from the ScCl precursor (**Fig. S3**).

Lanthanides with 4*f*- electrons, along with Y and Sc elements, share the same outer valence electron structure, resulting in similar reactivity and coordination chemistry. Indeed, a series of lanthanide MXenes (Ln$_2$CCl$_2$, Ln=Gd, Tb, Dy, Ho, Er, Lu) can been synthesized using the aforementioned carbon-intercalated 2D mono-halides strategy (**Figs. S4-S8** and **Table. S2**). Taking Er$_2$CCl$_2$ as an example, the XRD pattern and Rietveld refinement results (**Fig. 2a**) revealed that the crystal structure of the final product belongs to the space group *P*-3*m*1, which is consistent with the reported structures of Y$_2$CCl$_2$ and Sc$_2$CCl$_2$. This hexagonal symmetry was further corroborated by high-resolution transmission electron microscopy (HR-TEM) and selected area electron diffraction (SAED) patterns along the [000l] axis (**Fig. S9**). Additionally, a scanning electron microscopy (SEM) image of Er$_2$CCl$_2$ (**Fig. 2b**) displayed an accordion-like microstructure similar to that observed in multilayer MXenes produced by a top-down etching method[10], suggesting strong *vdW* interactions between the layers. Energy dispersive spectroscopy (EDS) analysis (**Fig. S10**) confirmed that the atomic ratio of Er to Cl is approximately 1:1, aligning with the expected stoichiometry. Further validation of the atomic arrangement was provided by scanning transmission electron microscopy (STEM) (**Fig. 2c**), which demonstrated that while the stacking model of layer of $Er_2CCl_2$ MXenes synthesized through such bottom-up approach differ from those derived from the MAX phase having twinned structure, the atomic arrangement in a single-layer remains the same[11]. X-ray photoelectron spectroscopy (XPS) results indicated that the oxidation state of Er is slightly less than +3, with a binding energy of 169.2 eV (**Fig. 2d**), positioned between that of erbium metal (167.6 eV)[15] and $ErCl_3$ (170.2 eV)[16]. Notably, the characteristic binding energy of Er-C in the C 1*s* orbital confirmed the intercalation of $C^{4-}$ into the positive layer $[Er_2Cl_2]^{4+}$ (**Fig. S11** and **Table S1**). Similarly, $Er_2CBr_2$ with Br terminal can be synthesized using Erbium mono-bromide (ErBr) as 2D building blocks (**Fig. S12**).

In addition to C, oxygen (O) and sulfur (S) atoms can diffuse into 2D building blocks. These atoms exhibit a tendency to form tetrahedral $[Ln_4O]$ and $[Ln_4S]$ basic units (**Figs. S13-S16**). Notably, light lanthanides—specifically from lanthanum (La) to europium (Eu)—are characterized by large atomic radii and high coordination numbers, as documented in the literature[17,18]. As a result, these elements do not readily form corresponding 2D MXenes structures. Instead, when these light lanthanide halogenides undergo electron transfer to carbon, they tend to transform into more stable three-dimensional crystal structures, exemplified by $La_3CBr_3$[19].

To highlight the critical importance of the composition of lanthanide chloride for MXenes formation, a series of $ErCl_n$ (n=1-3) compounds were synthesized *in situ* by adjusting the ratio of Er to $CuCl_2$, followed by reaction with graphite under identical conditions. The results clearly indicated that only ErCl, with 1:1 stoichiometry, successfully reacted with carbon to form $Er_2CCl_2$ MXenes. In contrast, when $ErCl_n$ (n=2-3) were used, the products contained unreacted graphite and erbium trichloride, highlighting the necessity of the 1:1 ratio for effective MXenes synthesis (**Fig. S17**). Given that erbium (Er) typically exhibits a +3 oxidation state, a lower stoichiometric ratio of erbium chloride can provide sufficient electrons to form the electride compound $[ErCl]^{2+}\cdot 2e^-$, which is advantageous for synthesizing corresponding MXene. A time-resolved *in situ* synchrotron radiation X-ray diffraction (SR-XRD) study provided detailed insights into the intermediate steps during the synthesis of lanthanide MXenes (**Figs. S18-S19**). It reveals that the phase evolution can be divided into two distinct steps (**Fig. 2e**). In Step I, occurring between 230 °C and 670 °C, the formation of *vdW* building block ErCl and the lattice expansion of the carbon are observed. This was evidenced by the emergence of a low-angle peak for ErCl (~9.43 °) and the shift of the characteristic carbon peak (~26.5 °) (**Fig. 2f**). In Step II, at higher temperatures, the disappearance of the carbon peak and the shift of the ErCl peak to a lower angle strongly supported the hypothesis that carbon is converted into $C^{4-}$ anions and subsequently intercalated into positively charged $[Er_2Cl_2]^{4+}$ layers.

**Semiconducting and magnetic properties**

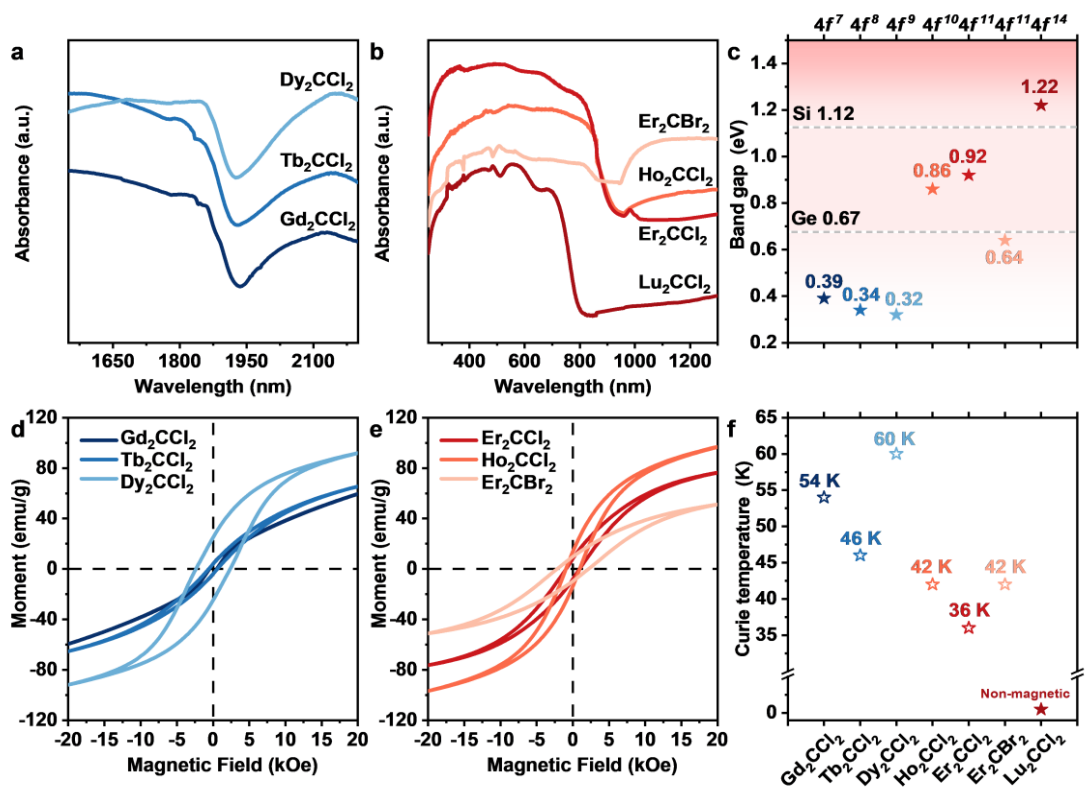

**Fig. 3 Semiconducting and magnetic properties characterization of Ln$_2$CT$_2$ (Ln=Gd, Tb, Dy, Ho, Er, Lu; T=Cl, Br)** (**a** and **b**) UV-vis-NIR diffuse reflectance spectroscopy of Ln$_2$CT$_2$ (Ln=Gd, Tb, Dy, Ho, Er, Lu; T=Cl, Br). (**c**) The indirect optical band gaps of the Ln$_2$CT$_2$. (**d** and **e**) Magnetic hysteresis loops of Ln$_2$CT$_2$ (Ln=Gd, Tb, Dy, Ho, Er; T=Cl, Br) at 2K. (**f**) Curie temperatures of Ln$_2$CT$_2$. The *x*-axis for **Fig. 3c** and **3f** is shared, with the lower *x*-axis representing Ln$_2$CT$_2$ arranged in ascending order of atomic number, and the upper *x*-axis representing the number of outer 4*f*-electrons of the lanthanide trivalent ions within the Ln$_2$CT$_2$.

Confining lanthanides, known for their unique *4f*-electrons, within 2D MXenes frameworks may impart these materials with diverse electronic and magnetic properties. Ultraviolet-visible-near infrared (UV-vis-NIR) spectra (**Fig. 3a** and **Fig. 3b**) showed that Ln$_2$CT$_2$ (Ln=Gd, Tb, Dy, Ho, Er, Lu; T=Cl, Br) presented an absorption edge characteristic of semiconductors in the near-infrared region. The Tauc method, based on the Kubelka-Munk theory, was applied to fit the spectral data, determining the optical indirect band gaps to span from 0.32 eV to 1.22 eV (**Fig. 3c** and **Fig. S20**). **Fig. 3c** depicts that Ln$_2$CCl$_2$ with the M site occupied by Gd, characterized by a half-filled *4f*-electron configuration, along with its neighboring elements Tb and Dy, all exhibit band gaps narrower than that of the commonly used semiconductor Ge (0.67 eV). Conversely, Ho$_2$CCl$_2$ and Er$_2$CCl$_2$, which are more aligned with Lu in terms of having a filled *4f*-electron configuration, manifest wider band gaps. Moreover, there is a trend of increasing band gap with an increase in the number of *4f*-electrons, peaking with the band gap of Lu$_2$CCl$_2$ exceeding that of Si (1.12 eV). Further modification of the terminal groups from Cl to the less electronegative Br resulted in a trend of decreasing band gap. These observations can be attributed to the influence of changes in the number of outer *4f*-electrons and terminal groups on orbital hybridization[20].

The magnetic hysteresis loops (**Fig. 4a** and **Fig. 4b**) demonstrated significant hysteresis in $Ln_2CT_2$ (Ln=Gd, Tb, Dy, Ho, Er; T=Cl, Br) MXenes, which attribute to the unpaired 4*f*-electrons at the M site. As the temperature increased, the curves progressively shifted to a paramagnetic state (**Fig. S21** and **Tables. S3-S8**), indicating that the synthesized $Ln_2CT_2$ MXenes exhibit ferromagnetic behavior. The Curie temperatures of samples, deduced from the thermal magnetization curves, were shown in **Fig. 3f** and **Fig. S22**, with temperatures ranging from 36 K to 60 K. Except for Lu, whose fully occupied 4*f* orbitals result in a relatively small magnetic moment, it is categorized as a paramagnetic element. The composition flexibility in lanthanide MXenes may further increase the Curie temperature and adjust the band structure[21,22].

**Semiconducting and magnetic mechanisms from first-principles calculations**

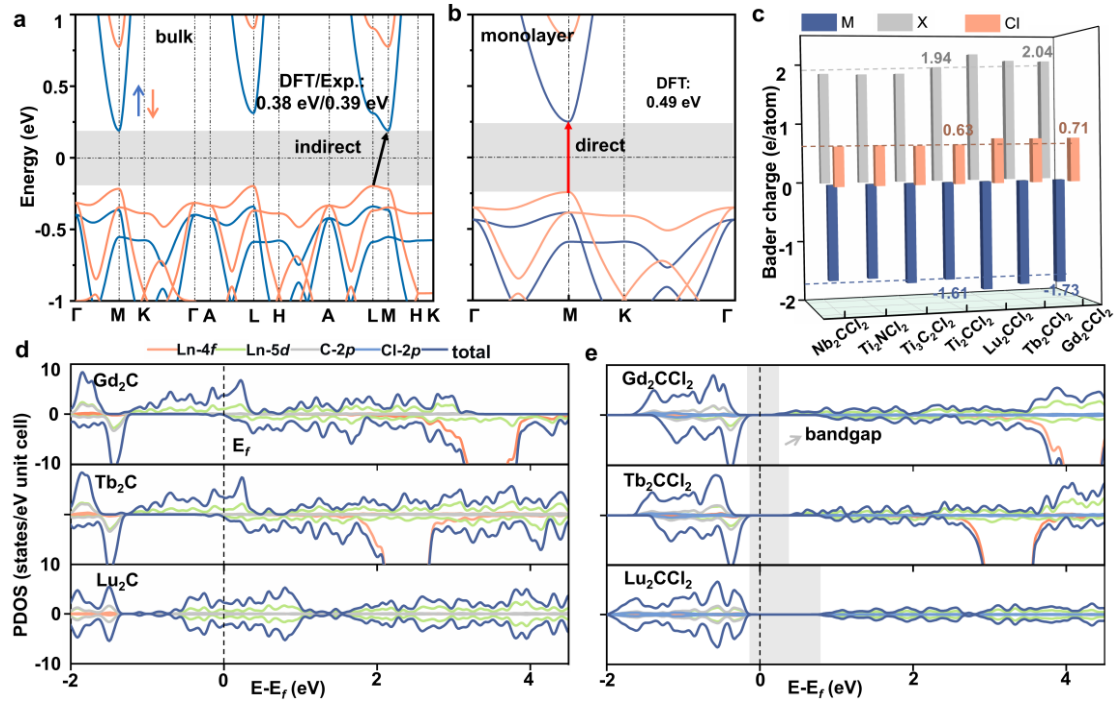

**Fig. 4 Electrical structures of the calculated $Ln_2CCl_2$ (Ln=Gd, Tb and Lu)**. (**a** and **b**) spin-resolved band structures of bulk and monolayer $Gd_2CCl_2$. (**c**) Comparison of Bader charge between the semiconductive $Ln_2CCl_2$ and some of the most common metallic MXenes with Cl termination. (**d** and **e**) Projected density of states (PDOS) of $Ln_2C$ and $Ln_2CCl_2$.

In order to understand the origin of semiconductive behavior in Lanthanide MXenes, density functional theory (DFT) calculations are performed on the $Gd_2CCl_2$, $Tb_2CCl_2$ and the nonmagnetic $Lu_2CCl_2$ with $4f^7$, $4f^8$ and $4f^{14}$ outer electrons (**Fig. S23**), respectively. The ferromagnetic ground state of the $Gd_2CCl_2$ is confirmed by the qSO method based on the generalized Bloch theorem[23,24] (**Fig. S24**). Similar with Si and Ge, all the studied $Ln_2CT_2$ presents indirect band nature, as the conduction band minimum (CBM, blue line) and the valence band maximum (VBM, red line) are not identical at momentum space, as shown in **Fig. 4a**. The calculated band gaps of $Gd_2CCl_2$, and $Lu_2CCl_2$, 0.38 eV and 1.01 eV (**Fig. 4a** and **Fig. S25**), are well consistent with our experimental value 0.39 eV and 1.22 eV, respectively. We further show that the monolayer $Gd_2CCl_2$ is a direct semiconductor with band gap of 0.54 eV, which is in stark contrast to its' bulk counterpart (**Fig. 4b**). Such a kind of dimension-dependent indirect to direct transition can also be seen in $MoS_2$ and other transition metal dichalcogenides due to the quantum confinement induces band shift in momentum space, aligning the CBM and VBM[25-27]. The band of $Tb_2CCl_2$ and $Lu_2CCl_2$ also show similar behaviors with $Gd_2CCl_2$ (**Fig. S26**). In direct-gap semiconductors, the electric dipole transition from VBM to CBM is allowed without the assistance of phono. The electron-hole pairs will recombine radiatively with a high efficiency, which is desirable for photoelectronic applications[28] and field-effect transistors[29].

The microscopic origin of the band gap states in MXenes is closely linked to the reduction of *d*-electron states around the Fermi energy level ($E_f$) and the electron withdrawal by surface terminals. All bare and the majority of terminated MXenes are metallic, due to the large *d*-electron states at the $E_f$ contributed by M element[30]. Here (**Fig. 4c**), as evidenced by the Bader charge analysis[31], the Cl functionalization shows more apparent effect on the electron redistribution in the $Ln_2CT_2$ than in the traditional MXenes, since more electrons are transferred from the Ln metal atoms to nonmagnetic X and Cl atoms. The PDOS (**Figs. 4d-e**) further indicates that the small number of 5*d*-electrons dominates the $E_f$ in bare $Ln_2C$, while the 4*f*-electrons are highly localized and stay away from the $E_f$ and nearly solely contribute to ferromagnetic spin splitting. These 5*d*-electrons are easily shifted to the conduction band entirely by surface Cl-terminal, causing band gaps in $Ln_2CT_2$. As a comparison, the surface Cl-group cannot completely exhaust the fertile *d*-electrons around the $E_f$ of the representative Ti- and Nb-based samples (**Fig. S27**), leaving non-zero electron states in the $E_f$, which, consequently, doesn't change the metallic behavior of traditional MXenes.

**Conclusions**

In this report, a carbon-intercalation of 2D *vdW* halides approach is developed to synthesize lanthanide MXenes, which, unlike traditional early transition metal MXenes, exhibit unique semiconducting and magnetic properties due to abundant inner *4f*-electrons. Theoretical calculations unveil that the spin splitting of lanthanide 4*f*-electrons and the depletion of the scarce lanthanide 5*d*-electrons near the Fermi level, induced by halogen terminals, are responsible for their ferromagnetic and semiconductive behavior, respectively. The integration of lanthanides significantly broadens the realm of 2D magnetic semiconductors with intrinsic ferromagnetism.

## Methods

### Materials

LnH$_2$ (Ln =Nd, Gd, Tb, Dy, Ho, Er and Lu) powders with purity of 99.9 wt.% and a mean particle size of ~75 μm were commercially obtained from Hunan Rare Earth Metal Materials Research Institute Co., Ltd. Scandium (Sc, 99.9 wt.%), yttrium (Y, 99.9 wt.%), lanthanum (La, 99.9 wt.%), copper oxide (CuO, 99.9 wt.%), sulfur (S, 99.5 wt%), copper chloride (CuCl$_2$, 98 wt.%), copper bromide (CuBr$_2$ 99 wt.%) and graphite (C, 99.5 wt.%), were purchased from Aladdin Chemical Reagent, China.

### Synthesis method of Y$_2$CCl$_2$, Sc$_2$CCl$_2$ and Ln$_2$CT$_2$ (Ln=Gd, Tb, Dy, Ho, Er, Lu; T=Cl, Br)

All the materials were carried out under Ar atmosphere in a glovebox and the obtained products are sensitive to moisture.

CuT$_2$(T=Br, Cl) powders, metal (Y, Sc, LnH$_2$) powders and graphene were used as the starting materials and weighed in a stoichiometric ratio of 1:2:1. First, the weighed graphite was putted into the quartz tube. CuT$_2$(T=Br, Cl) powders and metal powders were mixed, ground for 30 minutes, and then pressed into a 10 mm-diameter disk under approximately 1 MPa pressure using a mold. Subsequently, the disk was sealed with graphite inside a quartz tube under a vacuum of 10$^{-1}$ MPa. In order to ensure that the reaction proceeds in two steps, the samples arranged in accordance with the configuration depicted in **Fig. S1**. The quartz tube containing the sample was then placed in a muffle furnace and heated at a rate of 5 °C/min to the desired temperature for a reaction duration of 12 h. The experimental details were shown in supporting information (**Table S2**). After the reaction was completed, the sample was removed from the quartz tube and ground uniformly.

### Synthesis method of LaOCl, NdOCl ErOCl and ErSCl

All the materials were carried out under Ar atmosphere in a glovebox.

CuCl$_2$ powders, metal (ErH$_2$, La, NdH$_2$) powders and CuO or S powders were used as the starting materials and weighed in a stoichiometric ratio of 1:2:2. First, the weighed CuO or S powders was putted into the quartz tube. CuCl$_2$ powders and metal powders were mixed, ground for 30 minutes, and then pressed into a 10 mm-diameter disk under approximately 1 MPa pressure using a mold. The remaining synthetic steps were consistent with Ln$_2$CT$_2$. The experimental details were shown in supporting information (**Table S2**).

### Structural and morphological characterization methods

The structure of Y$_2$CCl$_2$, Sc$_2$CCl$_2$, Ln$_2$CT$_2$, ErOCl, ErSCl, LaOCl, NdOCl were characterized using an X-ray diffraction (XRD, D8 Advance, Bruker AXS, Germany) with Cu Kα radiation. XRD Rietveld refinement of samples was performed by TOPAS-Academic v6. Morphology and chemical composition was characterized by scanning electron microscopy (SEM, G300, ZEISS, Germany) equipped with energy dispersive X-ray spectroscopy (EDS) system and a transmission electron microscope (TEM, Talos, Thermo Fisher Scientific, America) with the embedded high-sensitivity Super-X detector. The bonding states were observed by X-ray photoelectron spectrometer (XPS, AXIS SUPERA+, Shimadzu, Japan). STEM image was collected on JEOL Grand ARM-300F operated at 300 keV. The camera length was 15 cm during the experiment process.

### Characterization measurements of structural evolution

CuCl$_2$ powders, ErH$_2$ powders and graphene were mixed in a stoichiometric ratio of 1:2:1. The mixed powder was then sealed in a quartz capillary tube with a diameter of ∼1.0 mm. The quartz capillary tube loaded samples were heated from 26 °C to 950 °C at a rate of 3 °C min$^{-1}$. The synchrotron radiation X-ray diffraction (SR-XRD) patterns during the synthesis process of Er$_2$CCl$_2$ were obtained at BL14B1 of the Shanghai Synchrotron Radiation Facility (SSRF) using X-ray with a wavelength of 0.6887 Å and calibrated using the LaB$_6$ standard from NIST (660b).

**Characterization of electronic structure and magnetic properties**

Ultraviolet-Visible-Near Infrared (UV-vis-NIR) diffuse reflectance spectroscopy of Ln$_2$CT$_2$ was performed using a Perkin Elmer Lambda 1050+. Samples were placed on a barium sulfate matrix and then flattened.

Magnetism of Ln$_2$CT$_2$ was characterized by a superconducting quantum interference device (SQUID-VSM).

**Acknowledgments**

We are grateful for funding support from the National Natural Science Foundation of China (U23A2093 and 12435017). We thank the Shanghai Synchrotron Radiation Facility (BL14B1) for the SR-XRD measurements. We acknowledge W. Wen for his guidance on SR-XRD testing.